\documentstyle[cite,epsf,11pt]{article}

\columnwidth\textwidth

\newcommand{\nn}{\nonumber}
\newcommand{\be}{\begin{equation}}
\newcommand{\ee}{\end{equation}}
\newcommand{\bea}{\begin{eqnarray}}
\newcommand{\eea}{\end{eqnarray}}
\newcommand{\ggpp}{\gamma\gamma \rightarrow \pi ^+\pi^-}

\newcommand{\gpmgpm}{\gamma\pi^{\pm} \rightarrow \gamma \pi^{\pm}}

\newcommand{\apmb}{\bar{\alpha}_\pi \pm \bar{\beta}_\pi}
\newcommand{\apb}{\bar{\alpha}_\pi + \bar{\beta}_\pi}
\newcommand{\amb}{\bar{\alpha}_\pi - \bar{\beta}_\pi}

\def\mytoday#1{{ } \ifcase\month \or
 January\or February\or March\or April\or May\or June\or July\or
 August\or September\or October\or November\or December\fi \space
 \number\year}

\begin{document}
\noindent
\begin{tabbing} \` BUTP--96/01\\ \end{tabbing}
\vspace*{1.5cm}

\begin{center}

{\Large{\bf{Charged Pion Polarizabilities to two Loops \footnote{Work
supported in part by Schweizerischer Nationalfonds}}}\\[1.5cm]}

Urs B\"urgi\\[1cm]

Institut f\"ur Theoretische Physik \\ Universit\"at Bern\\
Sidlerstrasse 5, CH-3012 Bern, Switzerland \\ [0.5cm] (e--mail:
buergi@butp.unibe.ch) \\[1.5cm]

\mytoday \\
\vspace{1cm}

{\bf{Abstract}}

\end{center}

\begin{quote}
{\small We evaluate the electric and magnetic
polarizabilities of  charged pions in the
framework of chiral perturbation
theory at next--to--leading order. This
 requires a  two--loop evaluation of the Compton amplitude
 near threshold.  We estimate the
 two new low--energy
constants which enter the chiral expansion at this order  with
resonance saturation. The  numerical results are compared with
 presently available experimental information. }
\end{quote}

\thispagestyle{empty}
\newpage

\renewcommand{\arraystretch}{1.5}

1. {\it Pion polarizabilities.} 
The first two terms in the threshold expansion
of the  Compton amplitude $\gpmgpm$
involve the electric charge and the electric ($\bar{\alpha}_\pi$) and
magnetic ($\bar{\beta}_\pi$) pion
polarizabilities.
To  predict the values of these, we invoke
 chiral perturbation theory (CHPT) \cite{WE79,GL84,GL85,LE94,chiral}. The resulting
 quark mass expansion of
$\bar{\alpha}_\pi$ and $\bar{\beta}_\pi$ is very similar to the one of
the
threshold parameters in $\pi\pi$ scattering. For illustration, we consider, in
addition to
the polarizabilities, the chiral expansion of the $I=0,$ $S$--wave scattering
length $a_0$. We have
\bea
& & \hspace{0.9cm}
\mbox{tree}\hspace{0.35cm}\mbox{1--loop}\hspace{0.45cm}\mbox{2--loops}\hspace{0.45cm
}\mbox{3--loops} \nn \\
a_0 & = &  \frac{7 \pi x}{2} \,
\left\{ \, 1 \; + \;\; A \, x \;\;\; + \;\; B \; x^2 \; +\;O(M_{\pi}^6) \; \right\} \nn \; , \\
\bar{\alpha}_\pi\pm\bar{\beta}_\pi& = &  \frac{\alpha}{M^3_{\pi} } \; \left\{ \; 0 \;
+ \;
A_{\pm}^{} \; x \;+\;B_{\pm}^{} x^2 \; + \; O(M^4_{\pi}) \; \right\} \nn \; , \\
x & = & \frac{M^2_{\pi}}{16 \pi^2 F^2_{\pi}} \; ,
\label{polchi1}
\eea
where $F_\pi=92.4$
MeV denotes the pion decay constant, and
 $\alpha=e^2/4\pi\simeq 1/137$.
 The first line indicates the number of
loops that are
needed to calculate the coefficients $A, A_\pm,\ldots$ .
 Whereas the
scattering length starts out with a tree graph contribution \cite{WE66}, the
leading order term in the polarizabilities is generated by one--loop diagrams
\cite{TE73,DH89,HO90}. The correponding numerical values are
\cite{TE73,DH89,HO90}
\footnote{We express the polarizabilities in units of $10^{-4}
\mbox{fm}^3$ throughout.}
\bea
\apmb =\left\{\begin{array}{l}
             0\\
             5.4\pm 0.8 \;\;.
              \end {array} \right.
\label{polchi2}
\eea
To estimate the reliability of this prediction, the corrections
 generated by two--loop diagrams  must be evaluated
\cite{LE92,BE94}. This has already been a\-chieved in the case of the
polarizabilities of the neutral pion \cite{BE94}.  In  this  letter we report
the results for  charged pions.
\vspace{0.3cm}

2. {\it Chiral expansion.} 
We first consider the threshold expansion of
the Compton amplitude 
in powers of the photon momentum. In the rest frame of the incoming pion, we
have
\be
 T = -2 \left[ {\bf{\epsilon}}_1 \cdot {\bf{\epsilon}}_2\:\!^*(e^2-4 \pi
M_{\pi}\bar{\alpha}_{\pi} |{\bf{q}}_1||{\bf {q_2}}|)-4 \pi M_{\pi}
\bar{\beta}_{\pi}({\bf{q}}_1\times{\bf{\epsilon}}_1)\cdot({\bf{q}}_2\times
{\bf{\epsilon}}_2\:\!^*) + \dots \right ] \; ,
\label{le1}
\ee
where ${\bf q}_i$ $({\bf {\epsilon}}_i)$ are the momenta (polarization
vectors)
of the  photons. The ellipsis denotes higher order terms in the photon momenta.
To arrive at the  expansion Eq. (\ref{polchi1}),
we consider
the Compton amplitude in the framework of CHPT,
 \be
T^{\mbox{\tiny{CHPT}}}=T_2+T_4+T_6+\cdots\;,
\label{le3}
\ee
where $T_{n}$ is of the order $p^{(n-2)}$.
This expansion  is most conveniently
performed in the framework of an effective lagrangian
\cite{WE79,GL84,GL85,LE94,chiral,LE92}. In the following we
ignore isospin breaking effects by
putting $m_u=m_d=\hat{m}$. The effective lagrangian is expressed in
terms of the pion field $U$, the quark mass matrix $\chi$ and the external
vector field $v_{\mu}$,
\be
{\cal L}_{\mbox{\tiny{eff}}} = {\cal L}_2 (U,\chi,v_{\mu})+\hbar \; {\cal L}_4 (U,\chi,v_{\mu})+ \hbar^2 \;
{\cal L}_6 (U,\chi,v_{\mu}) + \ldots \; ,
\label{le4}
\ee
where ${\cal L}_{n}$ denotes a term of order $p^{n}$.
The lagrangians ${\cal L}_2$ and ${\cal L}_4$ are given in Ref.
\cite{GL84}, whereas
the general structure of  ${\cal L}_6$  has recently been
determined in Ref. \cite{FS94}.
Given
${\cal L}_{\mbox{\tiny{eff}}} $, it is straightforward to expand the
S--matrix elements in powers of $\hbar$. This procedure automatically
generates the series (\ref{le3}).
 The expansion of the
polarizabilities is  obtained by comparing
 $T^{\mbox{\tiny{CHPT}}}$ with  the momentum expansion Eq. (\ref{le1}).
 The first term $T_2$ is
due to tree graphs and describes Compton scattering from a point--like
scalar particle. It does not contribute to the polarizabilities, because it
reduces to the first term in the expansion Eq. (\ref{le1}).
 $T_4$ is responsible for the leading order term
$A_\pm$ in (\ref{polchi1}). It contains two types of contributions: one--loop
graphs generated by
${\cal L}_2$ and tree graphs involving one vertex from ${\cal L}_4$. The
relevant momentum integrals have been performed in Ref. \cite{BC88} -- the
corresponding numerical result for the polarizabilities is displayed in Eq.
(\ref{polchi2}).
Finally, the
amplitude $T_6$  determines the next--to--leading corrections
$B_\pm$. It contains two--loop contributions
from the effective lagrangian ${\cal L}_2$ together with one--loop contributions
from ${\cal L}_2+{\cal L}_4$
and tree--level contributions from ${\cal L}_2+{\cal
L}_4+{\cal L}_6$.
We have found far more than 100
diagrams that contribute at this order. They  are
displayed in Refs. \cite{BU96b,BU96}.
\vspace{0.3cm}

3. {\it Next--to--leading order terms.}
We omit the details of the calculation that  is described in Refs.
\cite{BU96b,BU96}. Here we simply quote the results:
\bea
A_+&=&0\; ,\;\;\;A_-=\frac{2}{3}(\bar{l}_6-\bar{l}_5)\; , \nn \\
B_+ & = & 8 h^{r}_-(\mu) -\frac{4}{9}
                                 \left\{l(l+\frac{1}{2}\bar{l}_1+\frac{3}{2}\bar{l}_2)-\frac{53}{24}l
                                 + \frac{1}{2}\bar{l}_1
                                 +\frac{3}{2}\bar{l}_2+\frac{91}{72} +
                                 \Delta_+^{} \right\} \; ,
                                 \nn \\
 B_- & = & h^{r}_+(\mu)
                                 -\frac{4}{3}
                                 \left\{l(\bar{l}_1-\bar{l}_2+\bar{l}_6-\bar{l}_5-\frac{65}{12}) \right. \nn \\
& & \hspace{2.8cm} \left.
                                 - \frac{1}{3}\bar{l}_1
                                 -\frac{1}{3}\bar{l}_2+\frac{1}{4}\bar{l}_3-(\bar{l}_6-\bar{l}_5)\,\bar{l}_4
                                 +\frac{187}{108} + \Delta_-^{}
                                 \right\} \; , \nn \\ 
l & = & \ln \frac{M_{\pi}^2}{\mu^2} \; .
\eea
The quantities $\bar{l}_i$ denote low--energy constants from ${\cal L}_4$,
and $h_{\pm}^r(\mu)$ are  linear combinations of renormalized,
scale--dependent low--energy couplings from
 ${\cal L}_6$.
 [The scale $\mu$ is introduced by
dimensional regularization -- the result for the polarizabilities is, of
course, scale--independent.]
 Finally, the quantities $\Delta_\pm$ (generated partly by the two--loop
integrals displayed in Fig. 1) are pure numbers, independent of low--energy
constants and quark masses.

To obtain numerical results, we need the  values of
$\bar{l}_1,\ldots, \bar{l}_6-\bar{l}_5, h_+^{r}$ and  $h_-^{r}$.
 The constants $\bar{l}_i$ have been determined in Refs. \cite{GL84,BC94}. To obtain
an estimate of the couplings $h_\pm^r$, we use resonance saturation
 \cite{GL84,BE94,EG89,EP90} with vector --  and axialvector mesons
($J^{PC}=1^{--},1^{+-},1^{++}$) \cite{PE81,KA86,KO93,DH93,BA93}. We find
(the uncertainties in the couplings quoted here are more generous than the ones
given in Ref. \cite{BU96}) \bea h_+^{r}(M_{\rho}) & = & 0.3
\;\; \pm 2.0 \; , \nn \\ h_-^{r}(M_{\rho}) & = & 0.45 \pm 0.15\; .
\label{polchi5}
\eea

\begin{table}[bt]
\caption[Polarizability]{\label{polval}\small{Charged pion polarizability up to
two--loops in units of $10^{-4}\mbox{fm}^3$. The values in the brackets
are included in the two--loop result (column 6).}}\vspace{0.1cm}
\begin{center}
\begin{tabular}{|c||r|r|r|r||r||}  \hline
                           & ${\cal O}(E^{-1})$ &
\multicolumn{4}{|c|}{${\cal O}(E)$} \\
\cline{2-6}
                           & 1--loop & $h^{r}_{\pm}$ &
$\Delta_{\pm}$ & chiral logs & 2--loops \\ \hline $\apb$ &
0.00 & [0.15] & [0.16] & [0.04] & 0.31 \\ $\amb$ & 5.36 & [0.01] &
[0.50] & [-1.45] & -0.94 \\ $\bar{\alpha}_{\pi^{\pm}}$ & 2.68 & [0.08]
& [0.33] & [-0.70] & -0.31 \\ $\bar{\beta}_{\pi^{\pm}} $ & -2.68 &
[0.07] & [-0.17] & [0.75] & 0.63 \\ \hline
\end{tabular}
\end{center}
\end{table}
\vspace{0.3cm}

4. {\it Numerics.} 
The numerical values of the polarizabilities are shown in
Table 1 \footnote{ We use $F_{\pi}=92.4$ MeV,
$M_{\pi}=139.6$ MeV and $\bar{l}_1=-1.7\pm 1.0$, $\bar{l}_2= 6.1 \pm 0.5$,
$\bar{l}_3= 2.9 \pm 2.4$, $\bar{l}_4=4.3 \pm 0.9 $, $\bar{l}_6 - \bar{l}_5 = 2.7 \pm 0.4$.}. The one--loop
results are displayed in the second column, and the results of the two--loop
calculation are presented in the third to sixth columns.
\vspace{0.3cm}

{\it Comments:}
\begin{itemize}
\item[--]
The total two--loop contributions displayed in the sixth column amount to a
correction of $12\% (24\%)$ to the leading order value of $\bar{\alpha}_\pi
(\bar{\beta}_\pi)$. We therefore expect higher order corrections to be
substantially smaller.
\item[--]
The contribution from the low--energy constants $h_\pm^r$ is substantial in the
case of the sum ($\bar{\alpha}_\pi +\bar{\beta}_\pi)$, and smaller
otherwise. The contribution from $\rho$--exchange alone,
$(\bar{\alpha}_\pi+\bar{\beta}_\pi)_\rho=0.07$, is more than three times
smaller than the remaining two--loop contributions to this quantity.
  \item[--]
The fifth column  contains the contributions from the chiral logarithms,
i.e., the sum of the
 $\ln^2
M_{\pi}/\mu$ and $\ln M_{\pi}/\mu$ terms at the scale
$\mu=M_{\rho}$. As has been shown in Ref. \cite{WE79},
 the coefficients of the chiral {\it double
logarithms} in any Green function can be determined using renormalization group
arguments.
We refer the reader to Ref. \cite{BE94} for an illustration of the method in
the neutral pion case, and to Ref. \cite{CO95} for an application in elastic
$\pi\pi$ scattering. In the present case, the
method allows one to determine the coefficients $C_\pm$ in
\bea
{(\apmb)}_{2loops} & = &  C_{\pm} L_{\chi} + \ldots \; , \nn \\
L_{\chi} & = & \frac{\alpha M_{\pi}}{(16 \pi^2 F_{\pi}^2)^2} \ln
                           (\frac{M_{\pi}^2}{\mu^2} )\{ \ln
                           (\frac{M_{\pi}^2}{\mu^2}) +\frac{1}{2}
                           \bar{l}_1+\frac{3}{2}\bar{l}_2 \} \; ,
\label{polchi6}
\eea
with a one--loop calculation with ${\cal L}_2+{\cal L}_4$. This serves as an additional
check on the calculation. The result is $(C_+,C_-)=(-\frac{4}{9},0)$.
The contribution of this term to the polarizabilities at $\mu=M_\rho$ is
$(\bar{\alpha}_\pi,\bar{\beta}_\pi)=(0.16,0.16)$.

\begin{figure}
\unitlength1cm
\begin{picture}(1,1) \end{picture}
\epsfysize=3.5cm
\epsffile{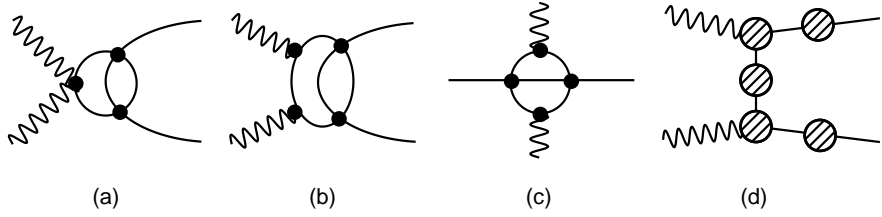}
\caption[Overlapping momentum integrals]{\label{overlapp}\small{Two--loop
graphs (vertex (a), box (b), acnode (c) and reducible diagrams (d)) generated
by ${\cal L}_2$ which contain overlapping loop momenta and contribute
to the Compton amplitude. A part of these contributions is included in
the quantities $\Delta_{\pm}$. The hatched circles
in diagram (d) denote self--energy and vertex  corrections at one--
and two--loop order. }}
 \end{figure}

\begin{table}[bt]
\caption[Polarizability of overlapping loops]{\label{polvallapp}\small{Logarithmic and total contributions
of the two--loop diagrams depicted in Fig. \ref{overlapp} to the polarizabilities
at the scale $\mu=M_{\rho}$ in units of $10^{-4}\mbox{fm}^3$.}}\vspace{0.1cm}
\begin{center}
\begin{tabular}{|c||c|c|c|c|c|c|c|c|c|}  \hline
                         & \multicolumn{3}{|c|}{ box (b)}
& \multicolumn{3}{|c|}{acnode (c)} & \multicolumn{3}{|c|}{reducible
diagrams (d)} \\
\cline{2-10}
                           & $\log$ & $\log^2$ & total & $\log$ &
$\log^2$ & total & $\log$ & $\log^2$ & total \\ \hline

$\apb$ & -0.04 & -0.45 & -0.37 & 0 & 0 & 0.02 & 0.1 & 0 & 0.08 \\
$\amb$ & -0.09 & -1.34 &-1.27 & -0.48 & -1.51 & -2.13 & -0.1 & 1.51 &
1.72 \\ \hline
\end{tabular}
\end{center}
\end{table}

\item[--]
We found it interesting to investigate the effect of the genuine two--loop
diagrams displayed in Fig. \ref{overlapp}. These contributions are
divergent at $d=4$.
 Here we consider the finite pieces of these
graphs, simply dropping the singular terms that occur in the renormalization
scheme
adopted in Ref. \cite{BU96}. The results are given in Table \ref{polvallapp},
where we display the values of the logarithmic terms proportional to $\ln^2
M_{\pi}/M_\rho$ and to $\ln M_{\pi}/M_\rho$, together with
the  total contribution. It is seen that the chiral double
logarithms
 are responsible for the large contribution to the
polarizabilities.
\item[--]
We have assumed that $h_\pm^r$ are saturated with resonance contributions at
the scale of the rho mass.
There is no
particular reason to prefer this scale to say $\mu=500$ MeV or $\mu=1$ GeV. The
corresponding results for the scales $\mu = 500, 700, 1000$ MeV are displayed
in Table \ref{polvalscale}. It
is seen that  the sum $\apb$ is nearly scale--independent,
 whereas
the difference $\amb$ changes by 20$\%$ by varying the scale from $\mu=500$ MeV
to $\mu$=1 GeV.
 \end{itemize}

{\it End of comments.}
\vspace{0.3cm}

\begin{table}[bt]
\caption[Polarizabilityscale]{\label{polvalscale}\small{Scale--dependence
of the polarizabilities in the  resonance saturation scheme.}}
\vspace{0.1cm}
\begin{center}
\begin{tabular}{|c||c|c|c|}  \hline
                           & $\mu=500$ MeV & $\mu=770$ MeV &
$\mu=1000$ MeV \\
\hline
$\apb$ & 0.31 & 0.31 & 0.30 \\ $\amb$ & 4.94 & 4.42 & 4.11 \\
$\bar{\alpha}_{\pi^{\pm}}$ & 2.63 & 2.37 & 2.20 \\
$\bar{\beta}_{\pi^{\pm}} $ & -2.32 & -2.06 & -1.90 \\ \hline
\end{tabular}
\end{center}
\end{table}

Our final result for the  polarizabilities of the charged pions at two--loop
accuracy is
\bea
\apb & = & \;\;\; 0.3 \pm 0.1 \hspace{1cm} (0.0) \nn \; , \\
\amb & = &  \;\;\; 4.4 \pm 1.0 \hspace{1cm} (5.4\pm 0.8)\nn \; , \\
\bar{\alpha}_{\pi^{\pm}} & = & \;\;\; 2.4 \pm 0.5 \hspace{1cm} (2.7 \pm 0.4)\nn \; , \\
\bar{\beta}_{\pi^{\pm}} & = & -2.1 \pm 0.5 \hspace{1cm} (-2.7 \pm 0.4)\; .
\label{polchi7}
\eea
The numbers in brackets denote the leading order result. The estimates of the
uncertainties stem  from the uncertainties
in the low--energy couplings $\bar{l}_1,\ldots,\bar{l}_6-\bar{l}_5,h_\pm^r$ and
do  contain neither effects from higher orders in the
quark mass expansion nor any correlations. 
(The errors are more generous than quoted in Ref. \cite{BU96}.) 
In addition, the relation
$(\bar{l}_6-\bar{l}_5)=6F_A/F_V$ \cite{TE73} has been used  [$F_A$
and $F_V$ denote axial and vector couplings in
$\pi\rightarrow\l\nu_l\gamma$.] This relation only holds at
leading order in the chiral expansion -- the implication of this fact for
the uncertainties in (\ref{polchi7}) has not yet been worked out.
 \vspace{0.3cm}

5. {\it Data versus CHPT.}
The Compton amplitude near threshold has been extracted from
 photon--nucleus scattering $\gamma p \rightarrow \gamma \pi^+ n$ \cite{AI86}
and from radiative pion nucleus scattering (Primakoff effect) $\pi^- Z
\rightarrow \pi^- \gamma Z$ \cite{AN84}.
 Analyzing the data with the constraint $\apb = 0$ gives
\be
\amb = \left\{ \begin{array}{r@{\: \pm \:} l} 40 & 24 \;\;\;\; \mbox{(Lebedev
                       \cite{AI86} )} \; ,\\ 13.6 & 2.8 \;\;\;
                       \mbox{(Serpukhov \cite{AN84} )} \; .\end{array} \right
                       .
\label{poldat1}
\ee
Relaxing the constraint
$\apb = 0$, the Serpukhov data  yield \cite{AN85}
\bea
\apb & = & \;\;1.4 \pm 3.1 \; \mbox{(stat.)} \; \pm 2.5 \; \mbox{(sys.)} \; ,
\nn \\ \amb & = & 15.6 \pm 6.4 \; \mbox{(stat.)} \; \pm 4.4 \; \mbox{(sys.)}
\; ,
\label{poldat2}
\eea
where we have evaluated  $\amb$ from
$\bar{\beta}_{\pi}$ and $\apb$ as given in Ref. \cite{AN85},
adding the errors in quadrature.

 The crossed amplitude $\ggpp$ can be measured
in pion--pair production in $e^+e^-$--collisions. At low energies, it is mainly
sensitive to $S$--wave scattering.
In Ref. \cite{KS93} unitarized S--wave
amplitudes have been constructed which contain $\amb$ as an adjustable
parameter. A fit to Mark II data at $E_{\pi\pi} < 800$ MeV \cite{mark} gives \cite{KS93}
\bea
\amb & = & \; 4.8 \pm 1.0 \; .
\label{poldat3}
\eea
(We have taken into account that the definition of the polarizability in
Refs. \cite{KS93,KP94,KP95} is $4\pi$ larger than the one used here).
The result
(\ref{poldat3}) contradicts the Serpukhov analysis Eq. (\ref{poldat1}). Taking
into account also D--waves, Kaloshin et al. \cite{KP94} find
\bea
\apb & = & \left\{ \begin{array}{r@{\: \pm \:} l} 0.22 & 0.06 \;\;\;\; \mbox{(
                       Mark II \cite{mark} )} \\ 0.30 & 0.04 \;\;\; \mbox{(
                       CELLO \cite{cello} ) } \; .  \end{array} \right .
\label{poldat4}
\eea
A detailed analysis of the same data has also been performed in
\cite{MP90}. The authors conclude
 that the
errors quoted in Eq. (\ref{poldat4}) are underestimated,
 see also Ref. \cite{PP94}. By reanalyzing the present data on the
angular distribution in $\ggpp$,
 Kaloshin et al. \cite{KP95} find
\bea
\apb & = & \left\{ \begin{array}{r@{\: \pm \:} l@{\: \pm \:} l} 0.22 & 0.07 \; \mbox{(stat.)}
                       & 0.04 \; \mbox{(sys.)} \;\;\;\; \mbox{(
                       Mark II \cite{mark} )} \\ 0.33 & 0.06 \; \mbox{(stat.)} & 0.01 \; \mbox{(sys.)} \;\;\;\; \mbox{(
                       CELLO \cite{cello} ) } \; .  \end{array} \right .
\label{poldat5}
\eea

Finally, we note that
the optical theorem relates the sum of the polarizabilities to the
total cross section $
\sigma^{\gamma\pi_{\pm}}_{tot}$. Writing an unsubtracted forward dispersion
relation
for one of the invariant amplitudes in $\gpmgpm$
gives
\be \bar{\alpha}_\pi+\bar{\beta}_\pi = \frac{M_{\pi}}{\pi^2}
\int_{4 M_{\pi}^2}^{\infty} \frac{ds'}{(s'-M_{\pi}^2)^2}
\sigma^{\gamma\pi}_{tot}(s') \; ,
\label{le2}
\ee
indicating that the sum of the polarizabilities has to be positive. Using a
model for the total cross section, Petrun'kin \cite{PE81} finds
\be
\apb = 0.39 \pm 0.04.
\label{petrun}
\ee
Note, however, that the lagrangian used \cite{PE81} to
estimate the cross section is not chiral invariant.
\vspace{0.3cm}

In order to clarify
the experimental situation,  new experiments to determine the pion
polarizabilities have been planned at Fermilab (E781 SELEX), Frascati (DA$\Phi$NE),
Grenoble (Graal facility) and at Mainz (MAMI). We
refer the reader to the section on hadron polarizabilities in Ref. \cite{BH95}
for details. \vspace{0.3cm}

We  now comment on the comparison of the chiral predictions for $\apmb$ with
the data.

\begin{itemize}

\item[--]

Our result $\amb = 4.4 \pm 1.0$ includes the leading and next--to--leading order terms.
It agrees within $1\frac{1}{2}$ standard deviation with the result $40 \pm 24$ found
at Lebedev \cite{AI86}. On the other hand, it is inconsistent with the value $13.6 \pm 2.8$
determined at Serpukhov \cite{AN84}.

\item[--]

The analysis done by Kaloshin et al. \cite{KS93,KP94,KP95} for $\apmb$
agrees within the error bars with the chiral predictions.

\item[--]

The value $\apb = 0.3 \pm 0.1$ includes the leading order term, generated by two--loop graphs.
It is positive, in agreement with the sum rule (\ref{le2}), and in good
agreement with the result (\ref{petrun}).

\end{itemize}

6. {\it Summary.} In summary, we have presented the expression for
the charged
pion polarizabilities at next--to--leading order in the chiral expansion.
In order to reduce the theoretical uncertainties,
it would be necessary to
determine the relevant low--energy constants with higher precision both, 
theoretically and experimentally. This
requires e.g. the evaluation of the chiral corrections to radiative beta decay
of the pion. Furthermore, it would be worthwhile to repeat
the analysis of Petrun'kin
\cite{PE81} in the framework of chiral symmetry.
On the experimental side, additional efforts are needed  to clarify the
situation.
\vspace{0.3cm}

I wish to thank
J\"urg Gasser for his help and advice throughout this work and for reading this
manuscript carefully.

\end{document}